\documentclass[reviewcopy]{elsarticle}

\usepackage[reviewcopy]{adndt}
\usepackage{longtable}

\usepackage{amsmath}

\usepackage{amssymb}

\biboptions{square,sort&compress}
\bibpunct[]{[}{]}{,}{n}{}{;}
\begin{document}

\begin{frontmatter}

\journal{Atomic Data and Nuclear Data Tables}


\title{Radiative rates for E1, E2, M1, and M2 transitions in Br-like ions with 43 $\le$ Z $\le$ 50}

  \author[One]{Kanti M. Aggarwal\fnref{}\corref{cor1}}
  \ead{K.Aggarwal@qub.ac.uk}

  \author[One]{Francis P. Keenan}


  \cortext[cor1]{Corresponding author.}

  \address[One]{Astrophysics Research Centre, School of Mathematics and Physics, Queen's University Belfast,\\Belfast BT7 1NN,
Northern Ireland, UK}


\date{16.12.2002} 

\begin{abstract}  
Energies and lifetimes are reported for the eight Br-like ions with 43 $\le$ Z $\le$ 50, namely Tc ~IX, Ru~X, Rh~XI, Pd~XII, Ag~XIII, Cd~XIV, In~XV, and Sn~XVI. Results are listed for the lowest 375 levels, which  mostly belong to the 4s$^2$4p$^5$, 4s$^2$4p$^4$4$\ell$, 4s4p$^6$,  4s$^2$4p$^4$5$\ell$,  4s$^2$4p$^3$4d$^2$,  4s4p$^5$4$\ell$, and  4s4p$^5$5$\ell$  configurations. Extensive configuration interaction among 39 configurations (generating 3990 levels) has been considered and the general-purpose relativistic atomic structure package ({\sc grasp}) has been adopted for the calculations. Radiative rates are listed  for all E1, E2, M1, and M2 transitions involving the lowest 375 levels. Previous experimental and theoretical energies are available for only a few levels of three, namely Ru~X, Rh~XI and Pd~XII. Differences with the measured energies are up to 4\% but the present results are an improvement (by up to 0.3 Ryd) in comparison to other recently reported theoretical data. Similarly for radiative rates and lifetimes, prior results are limited to those involving only  31 levels of the 4s$^2$4p$^5$, 4s$^2$4p$^4$4d, and 4s4p$^6$configurations  for the last four ions. Moreover, there are generally no discrepancies with our results, although the larger calculations reported here differ by up to two orders of magnitude for a few transitions. 

\vspace{0.5 cm}
\end{abstract}

\end{frontmatter}




\newpage

\tableofcontents
\listofDtables
\listofDfigures
\vskip5pc


\section{Introduction}

In a recent paper \cite{br42} we reported energy levels, lifetimes and  radiative decay rates (A-values) for five Br-like ions with 38 $\le$ Z $\le$ 42, i.e. Sr ~IV, Y~V, Zr~VI, Nb~VII, and Mo~VIII. Similar results for another important Br-like ion of tungsten (W~XL) have also  been published \cite{w40}. Here we extend the range of ions to those with 43 $\le$ Z $\le$ 50, i.e. Tc~IX, Ru~X, Rh~XI, Pd~XII, Ag~XIII, Cd~XIV, In~XV, and Sn~XVI. Ions of some of these elements, particularly Ag, Cd, In, and Sn, are important for the studies of laser-produced and fusion plasmas \cite{larry,denne}.

Laboratory measurements  for energy levels for these Br-like ions  are limited to only a few levels -- see section 2. Unfortunately, the theoretical situation is no better, although  A-values  \cite{bch}  for the  magnetic dipole (M1) and electric quadrupole (E2) transitions between the ground state levels (4s$^2$4p$^5$ $^2$P$^o_{3/2}$ -- $^2$P$^o_{1/2}$) of several Br-like ions, including those considered here, are available. These limited results are insufficient for modelling and diagnostics of  plasmas, particularly for fusion. The need for a large amount of atomic data and for a wide range of ions is particularly urgent with the developing ITER project. Therefore, in a recent paper Goyal  at al.  \cite{ag} have calculated  energies and lifetimes for the lowest 31 levels of  the 4s$^2$4p$^5$, 4s$^2$4p$^4$4d and 4s4p$^6$  configurations of the four Br-like ions  Ag~XIII, Cd~XIV, In~XV, and Sn~XVI.  For the calculations, they adopted the well known and widely used GRASP code  \cite{grasp0},  but included only limited CI (configuration interaction).  Similarly, they also listed A-values for electric  dipole (E1) transitions, but only from the ground state  4s$^2$4p$^5$ $^2$P$^o_{3/2,1/2}$ to higher-lying levels. Therefore, there is scope to extend significantly  their calculations. 

\section{Energy levels}

As in our earlier work \cite{br42,w40},  we have adopted the  fully relativistic multi-configuration Dirac-Fock (MCDF) atomic structure code developed by Grant  et al.  \cite{grasp0}. Since the code is  based on the $jj$ coupling scheme and includes higher-order relativistic corrections arising from the Breit interaction and QED (quantum electrodynamics) effects, it is suitable for the heavy ions considered here.  We note that this initial version of the MCDF code  has undergone many  revisions  and is now  known by the name  GRASP   \cite{grasp}. Again, as  in all  our work we employ the GRASP0 version (available at  {\tt http://web.am.qub.ac.uk/DARC/}), which has been revised by one of its authors (Dr. P. H. Norrington).  As expected, the results obtained are comparable with those from other versions, such as GRASP2K \cite{grasp2k, grasp2kk}.

Extensive configuration interaction  (CI)  has been incorporated in our calculations of energy levels  and A-values. These calculations include the same 39 configurations as in \cite{br42} and are:  4s$^2$4p$^5$, 4s$^2$4p$^4$4d/4f, 4s4p$^6$, 4p$^6$4d/4f, 4s4p$^5$4d/4f, 4s$^2$4p$^3$4f$^2$/4d$^2$/4d4f, 4s$^2$4p$^2$4d$^3$, 4s$^2$4p4d$^4$,  4s$^2$4p$^2$4d$^2$4f, 4s4p$^3$4d$^3$, 4p$^5$4d$^2$, 3d$^9$4s$^2$4p$^5$4d/4f, 3d$^9$4s$^2$4p$^6$, 4s4p$^5$5$\ell$,   4p$^6$5$\ell$,  4s$^2$4p$^4$5$\ell$, and 3d$^9$4s$^2$4p$^5$5$\ell$. As in our work for other Br-like ions \cite{br42}, these configurations have been carefully chosen on the basis of their interacting energy ranges. In total   3990 levels are generated, but for brevity we only consider  the lowest 375 levels for each ion. Furthermore,   we have adopted the option of  `extended average level' (EAL),  in which a weighted (proportional to 2$j$+1) trace of the Hamiltonian matrix is minimised. The code has several other choices for optimisation, such as average level (AL) and extended optimal level (EOL), but EAL has been preferred. This is because the results obtained with the AL option are comparable with those of EAL as already discussed and demonstrated by us for several other ions, such as  of Kr \cite{kr} and Xe \cite{xe}. Similarly, the EOL option may provide a slightly more accurate data for a few defined levels, and is useful if the experimental energies are known, which is not the case for the presently studied ions.

In Tables 1--8 we list our energetically lowest 375  levels for the eight  Br-like ions with 43 $\le$ Z $\le$ 50. Before we discuss results, we note that the $LSJ$ labels assigned  to the levels in Tables 1--8  are only for guidance, because their identification is not always unique, as  levels from different configurations mix strongly and for a few  the eigenvector of the same level/configuration dominates for two (or more) levels. This  problem is frequently encountered in all atomic structure calculations, particularly when the CI is important, as  for all Br-like ions \cite{br42, w40}. Fortunately, the ambiguity in the configuration/level ($LSJ$) designation is only for a few levels in each ion. For illustration, in Table A we list the mixing coefficients for the lowest 50 levels of  Sn~XVI.  It is clear that levels such as 5/16, 7/13, 10/25, 20/23, and 44/47 are highly mixed, and several eigenvectors have comparable magnitude. Nevertheless,  the associated $J^{\pi}$ values provided in Tables 1--8 are definitive for all levels and ions, but the $LSJ$ designations may (inter)change depending on the calculations (with differing amount of CI) and/or  the codes employed and indeed author preferences.

In Table B we compare energies for the  levels of Ru~X which are common between our calculations and the measurements of Even-Zohar and Fraenkel \cite{even}. Although there are only 9 levels, the differences are up to $\sim$4\%. For the 4s$^2$4p$^5$ $^2$P$^o_{1/2}$ level (2) our energy is lower but is higher for the rest. It is interesting to note that the Breit and QED contributions for level 2 are significant, i.e. --0.00482 and 0.00025 Ryd, respectively, but for others are not. As a result, our Coulomb energy for level 2 is 0.30050 Ryd, much closer to the measurement. Even-Zohar and Fraenkel \cite{even} have also measured energies for a few levels of Rh~XI and Pd~XII, and these are compared with our results in Tables C and D, respectively. The discrepancies for these levels are similar to those found for Ru~X. We also note that the effect of Breit and QED corrections is most dominant on the ground level energy. As an example, the Breit and QED contributions on the ground level energy of Sn~XVI are 8.74 and 5.50 Ryd, respectively. However, among higher excited levels their additive contributions are only up to $-$0.08 Ryd.

Also included in Tables B, C and D are our results obtained with the {\em Flexible Atomic Code} ({\sc fac}) of Gu \cite{fac}, which is fully relativistic and is available from the website {\tt http://sprg.ssl.berkeley.edu/$\sim$mfgu/fac/}. As for other Br-like ions \cite{br42}, we have performed a series of calculations with this code, and our largest calculation includes 12~137 levels, the additional ones arising from the  4p$^6$6$\ell$,  4s4p$^5$6$\ell$, 4s$^2$4p$^4$6$\ell$,   4p$^6$7$\ell$,  4s4p$^5$7$\ell$, 4s$^2$4p$^4$7$\ell$, 4s$^2$4p$^3$5$\ell^2$, 4s4p$^4$5$\ell^2$, 3p$^5$3d$^{10}$4s$^2$4p$^6$, 3p$^5$3d$^{10}$4s$^2$4p$^5$4d, and 3p$^5$3d$^{10}$4s$^2$4p$^5$4f configurations. The main aim of this exercise is to assess the impact of additional CI on the determination of energy levels (and the A- values, see section 3). However,  there are no appreciable discrepancies between these results and those with {\sc grasp}, although we do observe some variations in level orderings. This is to be expected because both codes are fully relativistic and a similarity of  results has already been noted for a range of ions, including Br-like \cite{br42}. Therefore, our conclusion remains the same as earlier \cite{br42, w40} that this large expansion of up to 12~137 levels  is not helpful in improving the energy levels, although extremely  large calculations involving over a million levels (or configuration state functions, CSF) may improve the accuracy  as  shown by Froese Fischer \cite{cff} and Bogdanovich et al. \cite{bog}  for W~XL. Unfortunately, our computational resources do not allow us to perform such large calculations. Additionally, for some Br-like ions, such as Sr~IV and W~XL, measurements are available for many more levels and therefore it becomes comparatively easier to assess (and/or to improve) the accuracy of calculations, but not for the ions considered here.

Finally, we note that there is no discrepancy for any level and ion with the recent results of Goyal et al. \cite{ag}, who have performed similar calculations with the same version of the {\sc grasp} code, but with CI among only eight configurations, namely 4s$^2$4p$^5$, 4s$^2$4p$^4$4d/4f, 4s4p$^6$,  4s4p$^5$4d/4f, and 4s$^2$4p$^3$4d$^2$/4f$^2$.  This is in total contrast with their earlier calculations \cite{sam} for five  other Br-like ions with 38 $\le$ Z $\le$ 42, which cannot be reproduced as already discussed by us \cite{ak}. However, the above eight  configurations generate 470 levels in total but \cite{ag} reported energies for only 31 levels of the 4s$^2$4p$^5$, 4s$^2$4p$^4$4d and 4s4p$^6$  configurations. This limited expansion of the wave functions reduces the accuracy of the calculated data, as demonstrated  by us \cite{ak}. Among the present ions of interest we compare our results with their calculations for Sn~XVI alone, in Table E. Due to the limited CI included by Goyal et al. \cite{ag}, their calculated energies are higher by up to $\sim$0.3 Ryd, almost the same amount as previously noted for other Br-like ions \cite{ak}.

\section{Radiative rates}\label{sec.eqs}

The absorption oscillator strength ($f_{ij}$, dimensionless)  for all types of  transition ($i \to j$)  and the radiative decay rate A$_{ji}$ (in s$^{-1}$) are connected by the following expression:

\begin{equation}
f_{ij} = \frac{mc}{8{\pi}^2{e^2}}{\lambda^2_{ji}} \frac{{\omega}_j}{{\omega}_i}A_{ji}
 = 1.49 \times 10^{-16} \lambda^2_{ji} \frac{{\omega}_j}{{\omega}_i} A_{ji}
\end{equation}
where $m$ and $e$ are the electron mass and charge, respectively, $c$  the velocity of light,  $\lambda_{ji}$  the transition wavelength in $\rm \AA$, and $\omega_i$ and $\omega_j$  the statistical weights of the lower $i$ and upper $j$ levels, respectively. Similarly, the oscillator strength $f_{ij}$, A-values  and the line strength $S$ (in atomic units, 1 a.u. = 6.460$\times$10$^{-36}$ cm$^2$ esu$^2$) are related by the  standard equations given in \cite{w40}. 

In Tables 9--16 we present results  for transitions in the eight Br-like ions with 43 $\le$ Z $\le$ 50, from the lowest three to higher excited levels, with the full tables  available online in the electronic version.  Included in these tables are the  transition (energies) wavelengths ($\lambda_{ij}$ in ${\rm \AA}$), radiative rates (A$_{ji}$ in s$^{-1}$), oscillator strengths ($f_{ij}$, dimensionless), and line strengths ($S$ in a.u.) for $\sim$18~000 E1 transitions among the lowest 375 levels.  The listed wavelengths (and other parameters) are based on the Breit and QED-corrected theoretical energies,  given in Tables 1--8, where the {\em indices} used to represent the lower and upper levels of a transition are also defined. Corresponding  A-values for $\sim$27~000 E2, $\sim$19~000  M1, and $\sim$25~000 M2 transitions are also included in these tables, as well as the ratio of their velocity (Coulomb gauge) and length (Babushkin gauge) forms, but only for  E1 transitions.  We also note that (if required)  the corresponding results for f- or S-values can be  obtained using Eqs. (1-5) given in \cite{w40}, and the number of transitions are not the same for all ions, because their level orderings are not common. 

The only data available for comparison purposes are those  of Goyal et al. \cite{ag}, where there are no discrepancies for calculations performed with the {\em same} CI. However, significant differences are noted with the present calculations with larger CI, i.e. among 3990 levels described in section 2. For illustration, we compare the two sets of f-values in Table F for  transitions in Sn~XVI. While there are differences of 50\% for several transitions, for a few the discrepancies are larger -- see for example, 1--27, 1--30 and 2-13.  The latter two are comparatively weak transitions, but 1--27 is not and the f-values differ by almost a factor of three. Interestingly the ratio (R) of the velocity and length forms in both calculations is 0.88, and hence does not provide any clue to the accuracy estimate. However, this is not surprising as similar examples have been noted many times in the past. On the other hand, inclusion of larger CI in a calculation generally (but not necessarily) improves the accuracy of the calculated radiative data, and a clear example of this has already been discussed for the energy levels in section 2. For f-values we include in Table F our results  obtained from the {\sc fac} code with 3990 level calculations (FACa), as with {\sc grasp}, as well as a larger one with 12~137 levels (FACb). The two sets of f-values from GRASP and FACa agree within $\sim$20\% for most transitions, although differences for a few weaker ones (such as 2--13/26) are up to 30\%. Similarly for some isolated (but weak) transitions (such as 1--30), differences are up to a factor of two. Such differences in f-values are common among calculations with different codes, because of the methodologies of solving the equations and/or the choice(s) of the local central potential. Nevertheless, it is heartening to note that the effect of additional CI included in FACb is of little consequence as far as the transitions of Table F are concerned. The  comparison shown in Table F for a few transitions of Sn~XVI confirms, once again, that the CI included in out GRASP calculations is sufficient to produce accurate results for radiative rates.

D'Arcy et al. \cite{darcy} have reported f- values for a few transitions of Sn~XVI, but have provided little information about the calculations. In Table G we compare our results with theirs. Differences for most transitions are less than 50\%, but are larger for two, namely 17--234 (4s$^2$4p$^4$4d~$^4$F$_{7/2}$ -- 4s$^2$4p$^4$5p~$^4$P$_{5/2}$) and 24--247 (4s$^2$4p$^4$4d~$^2$F$_{7/2}$ -- 4s$^2$4p$^4$5p~$^4$D$_{5/2}$). Since this comparison is limited to only 13 transitions, all of which are generally weak,  it is difficult to place an accuracy estimate on either of the two calculations. We discuss these further below.

A general criterion to assess the accuracy of radiative rates is to compare the velocity and length forms of A- (f-) values, i.e. R should be close to unity. However, often this is not the case even for strong dipole allowed transitions, as already noted above for 1--27. Nevertheless, we provide some statistics for the E1 transitions of Sn~XVI, listed in Table~16.  There are 1436 out of 17~272  transitions for which f(E1) $\ge$ 0.01, and about a third differ by over 20\%. However, only 165 have R $\ge$ 50\% but  for 7 the ratio is up to two orders of magnitude (26--82 f=0.03, 28--41 f = 0.01, 29--42 f = 0.01, 29--82 f = 0.04, 30--60 f = 0.04, 31--79 f = 0.03, and 139--259 f = 0.01), i.e. none of these is too strong.  For weaker transitions the ratio R is up to several orders of magnitude for a few, but similar to those noted for other Br-like ions \cite{br42,w40}. In conclusion, we may state that for a majority of strong transitions the A-values are accurate to $\sim$20\%, but for some the accuracy is lower. A better assessment of the accuracy of our reported data may perhaps be performed with other calculations, which may be available in future. Finally,   there are no discrepancies with the Bi{\' e}mont et al.  \cite{bch} A-values for the M1 and E2 (4s$^2$4p$^5$) $^2$P$^o_{3/2}$ -- $^2$P$^o_{1/2}$ transitions.

\section{Lifetimes}

The lifetime $\tau$ of a level $j$ is determined as 1.0/$\Sigma_{i}$A$_{ji}$. Since this is a measurable quantity it helps in assessing the accuracy of A-values, particularly when a single (type of) transition dominates. Unfortunately, no measurements of $\tau$ are available for the levels of Br-like ions, but in Tables 1--8  we list our calculated results  for the lowest 375.  The calculations of $\tau$  include A-values from all types of transitions, i.e. E1, E2, M1, and M2. 

\section{Conclusions}

Energy levels and radiative rates (for E1, E2, M1, and M2 transitions) are reported  for the lowest 375 levels of  eight Br-like ions with 43 $\le$ Z $\le$ 50, for which the {\sc grasp}  code has been adopted. Lifetimes for these levels are also listed although no other comparable theoretical data or measurements are currently available in the literature. Based on comparisons with the limited measurements our energy levels are assessed  to be accurate to better than 4\%, for all ions.  However, scope remains for improvement. A similar assessment of accuracy for the corresponding A-values is not feasible, mainly because of the paucity of other comparable  results. However, for strong transitions (with large f- values),   the accuracy for A-values and  lifetimes may be $\sim$20\%. 



\ack
KMA  is thankful to  AWE Aldermaston for financial support. 

\begin{appendix}

\def\thesection{} 

\section{Appendix A. Supplementary data}

Owing to space limitations, only parts of Tables 9--16  are presented here, the full tables being made available as supplemental material in conjunction with the electronic
publication of this work. Supplementary data associated with this article can be found, in the online version, at doi:nn.nnnn/j.adt.2015.nn.nnn.

\end{appendix}



\clearpage
\newpage


\renewcommand{\baselinestretch}{1.0}
\footnotesize
\begin{longtable}{@{\extracolsep\fill}rllr@{}}
\caption{Mixing coefficients (MC) for the lowest 50 levels of  Sn~XVI.  Numbers outside and inside a bracket correspond to MC and the level, respectively. See Table 1 for definition of all levels.}
\\
Index  & Configuration           & Level              &  Mixing coefficients     \\
 \hline\\
\endfirsthead\\
\caption[]{(continued)}
Index  & Configuration           & Level              &  Mixing coefficients   \\
\hline\\
\endhead
     1   &   4s$^2$4p$^5$		       &   $^2$P$^o_{3/2 }$    &    0.99(  1)					    \\
    2   &   4s$^2$4p$^5$		       &   $^2$P$^o_{1/2 }$    &    0.99(  2)					    \\
    3   &   4s4p$^6$			       &   $^2$S$  _{1/2 }$    &    0.87(  3)$-$0.48( 26) 			    \\
    4   &   4s$^2$4p$^4$($^3$P)4d	       &   $^4$D$  _{5/2 }$    &   $-$0.27( 12)$+$0.84(  4)$-$0.24( 15)$-$0.27( 23)    \\
    5   &   4s$^2$4p$^4$($^3$P)4d	       &   $^4$D$  _{3/2 }$    &   $-$0.77(  5)$+$0.34( 19)$-$0.23( 31)$+$0.33( 16)$-$0.20( 25)			    \\
    6   &   4s$^2$4p$^4$($^3$P)4d	       &   $^4$D$  _{7/2 }$    &   $-$0.46( 17)$+$0.82(  6)$+$0.28( 24)		    \\
    7   &   4s$^2$4p$^4$($^3$P)4d	       &   $^4$D$  _{1/2 }$    &    0.66(  7)$-$0.39( 11)$-$0.41( 30)$+$0.48( 13)    \\
    8   &   4s$^2$4p$^4$($^3$P)4d	       &   $^4$F$  _{9/2 }$    &    0.88(  8)$+$0.45( 21) 			    \\
\hline  											       								   					      
\end{longtable}

                               
\begin{flushleft}                                                                               
                               
{\small
                                                                               
}                                                                                               
                               
\end{flushleft} 

\renewcommand{\baselinestretch}{1.0}
\footnotesize
\begin{longtable}{@{\extracolsep\fill}rlllllr@{}}
\caption{Comparison of energies (Ryd) for some  levels of Ru~X. }
Index$^a$  & Configuration         & Level      & EXP.    & GRASP  &   FAC  & \%$\Delta$E  (EXP.-- GRASP)  \\  \\
 \hline\\
\endfirsthead\\
\caption[]{(continued)}
Index$^a$  & Configuration         & Level      & EXP.    & GRASP  &   FAC  & \%$\Delta$E  (EXP.-- GRASP)  \\  \\
\hline\\
\endhead 
  1  & 4s$^2$4p$^5$	&  $^2$P$^o_{3/2}$  & 0.00000  & 0.00000 & 0.00000 &  0.0 \\
  2  & 4s$^2$4p$^5$	&  $^2$P$^o_{1/2}$  & 0.30122  & 0.29593 & 0.29791 &  1.8 \\
 48  & 4s$^2$4p$^4$5s	&  $^4$P$_{5/2}$    & 6.35869  & 6.57994 & 6.58739 & $-$3.5 \\
 51  & 4s$^2$4p$^4$5s	&  $^4$P$_{3/2}$    & 6.42122  & 6.66128 & 6.66294 & $-$3.7 \\
 57  & 4s$^2$4p$^4$5s	&  $^4$P$_{1/2}$    & 6.55883  & 6.77825 & 6.78696 & $-$3.3 \\
 61  & 4s$^2$4p$^4$5s	&  $^2$P$_{3/2}$    & 6.64029  & 6.86595 & 6.87057 & $-$3.4 \\
 66  & 4s$^2$4p$^4$5s	&  $^2$P$_{1/2}$    & 6.69557  & 6.93747 & 6.93680 & $-$3.6 \\
 73  & 4s$^2$4p$^4$5s	&  $^2$D$_{5/2}$    & 6.79679  & 7.05426 & 7.05550 & $-$3.8 \\
 74  & 4s$^2$4p$^4$5s	&  $^2$D$_{3/2}$    & 6.80872  & 7.06859 & 7.06873 &$-$3.8 \\
110  & 4s$^2$4p$^4$5s	&  $^2$S$_{1/2}$    &          & 7.44119 & 7.45107 &      \\
\\  \hline            								                	 
\end{longtable}   								   					       
			      							   					       

														       
\begin{flushleft}													       
{\small
$a$: see Table 2 \\
EXP: Even-Zohar and Fraenkel \cite{even} \\
GRASP: present calculations from the {\sc grasp} code with 3990 levels \\ 
FAC: present calculations from the {\sc fac} code with 12~137 levels \\ 		       
}															       
\end{flushleft} 


\renewcommand{\baselinestretch}{1.0}
\footnotesize
\begin{longtable}{@{\extracolsep\fill}rlllllr@{}}
\caption{Comparison of energies (Ryd) for some  levels of Rh~XI. }
Index$^a$  & Configuration         & Level      & EXP.    & GRASP  &   FAC  & \%$\Delta$E  (EXP.-- GRASP)  \\  \\
 \hline\\
\endfirsthead\\
\caption[]{(continued)}
Index$^a$  & Configuration         & Level      & EXP.    & GRASP  &   FAC  & \%$\Delta$E  (EXP.-- GRASP)  \\  \\
\hline\\
\endhead 
  1  & 4s$^2$4p$^5$	&  $^2$P$^o_{3/2}$  & 0.00000  & 0.00000 & 0.00000 &  0.0 \\
  2  & 4s$^2$4p$^5$	&  $^2$P$^o_{1/2}$  & 0.35357  & 0.34817 & 0.35017 &  1.5 \\
 61  & 4s$^2$4p$^4$5s	&  $^4$P$_{5/2}$    & 7.22918  & 7.46403 & 7.47319 & $-$3.2 \\
 64  & 4s$^2$4p$^4$5s	&  $^4$P$_{3/2}$    & 7.29592  & 7.55072 & 7.55375 & $-$3.5 \\
 73  & 4s$^2$4p$^4$5s	&  $^4$P$_{1/2}$    & 7.45072  & 7.68338 & 7.69325 & $-$3.1 \\
 83  & 4s$^2$4p$^4$5s	&  $^2$P$_{3/2}$    & 7.56012  & 7.79844 & 7.80497 & $-$3.2 \\
 90  & 4s$^2$4p$^4$5s	&  $^2$P$_{1/2}$    & 7.61710  & 7.87280 & 7.87376 & $-$3.4 \\
102  & 4s$^2$4p$^4$5s	&  $^2$D$_{5/2}$    & 7.72261  & 7.99444 & 7.99775 & $-$3.5 \\
107  & 4s$^2$4p$^4$5s	&  $^2$D$_{3/2}$    &          & 8.01131 & 8.01333 &	  \\
142  & 4s$^2$4p$^4$5s	&  $^2$S$_{1/2}$    &          & 8.43279 & 8.44359 &	  \\
\\  \hline            								                	 
\end{longtable}   								   					       
			      							   					       

														       
\begin{flushleft}													       
{\small
$a$: see Table 3 \\
EXP: Even-Zohar and Fraenkel \cite{even} \\
GRASP: present calculations from the {\sc grasp} code with 3990 levels\\ 
FAC: present calculations from the {\sc fac} code with 12~137 levels \\ 		       
}															       
\end{flushleft} 

\clearpage
\newpage

\renewcommand{\baselinestretch}{1.0}
\footnotesize
\begin{longtable}{@{\extracolsep\fill}rlllllr@{}}
\caption{Comparison of energies (Ryd) for some  levels of Pd~XII. }
Index$^a$  & Configuration         & Level      & EXP.    & GRASP  &   FAC  & \%$\Delta$E  (EXP.-- GRASP)  \\  \\
 \hline\\
\endfirsthead\\
\caption[]{(continued)}
Index$^a$  & Configuration         & Level      & EXP.    & GRASP  &   FAC  & \%$\Delta$E  (EXP.-- GRASP)  \\  \\
\hline\\
\endhead 
  1  & 4s$^2$4p$^5$	&  $^2$P$^o_{3/2}$  & 0.00000  &   0.00000  &  0.00000   &  0.0 \\
  2  & 4s$^2$4p$^5$	&  $^2$P$^o_{1/2}$  & 0.41171  &   0.40632  &  0.40838   &  1.3 \\
 79  & 4s$^2$4p$^4$5s	&  $^4$P$_{5/2}$    & 8.14287  &   8.39061  &  8.40099   & $-$3.0 \\
 88  & 4s$^2$4p$^4$5s	&  $^4$P$_{3/2}$    & 8.21365  &   8.48226  &  8.48226   & $-$3.3 \\
102  & 4s$^2$4p$^4$5s	&  $^4$P$_{1/2}$    & 8.38480  &   8.63040  &  8.64123   & $-$2.9 \\
109  & 4s$^2$4p$^4$5s	&  $^2$P$_{3/2}$    & 8.52899  &   8.77940  &  8.78735   & $-$2.9 \\
118  & 4s$^2$4p$^4$5s	&  $^2$P$_{1/2}$    & 8.58824  &   8.85661  &  8.85878   & $-$3.1 \\
130  & 4s$^2$4p$^4$5s	&  $^2$D$_{5/2}$    & 8.69781  &   8.98289  &  8.98777   & $-$3.3 \\
133  & 4s$^2$4p$^4$5s	&  $^2$D$_{3/2}$    & 8.71493  &   9.00237  &  9.00578   & $-$3.3 \\
176  & 4s$^2$4p$^4$5s	&  $^2$S$_{1/2}$    &	       &   9.47897  &  9.49039   &	\\
\\  \hline            								                	 
\end{longtable}   								   					       
			      							   					       

														       
\begin{flushleft}													       
{\small
$a$: see Table 4 \\
EXP: Even-Zohar and Fraenkel \cite{even} \\
GRASP: present calculations from the {\sc grasp} code with 3990 levels\\ 
FAC: present calculations from the {\sc fac} code with 12~137 levels \\ 		       
}															       
\end{flushleft} 


\renewcommand{\baselinestretch}{1.0}
\footnotesize
\begin{longtable}{@{\extracolsep\fill}rllll@{}}
\caption{Comparison of energies (Ryd) for the lowest 31 levels of Sn~XVI.} \\
Index  & Configuration         & Level               & GRASPa & GRASPb  \\
 \hline\\
\endfirsthead\\
\caption[]{(continued)}
Index  & Configuration         & Level               & GRASPa & GRASPb  \\
\hline\\
\endhead 
  1  & 4s$^2$4p$^5$	       &  $^2$P$^o_{3/2}$    & 0.0000 & 0.0000  \\
  2  & 4s$^2$4p$^5$	       &  $^2$P$^o_{1/2}$    & 0.7069 & 0.7070  \\
  3  & 4s4p$^6$ 	       &  $^2$S$  _{1/2}$    & 4.0812 & 4.1143  \\
  4  & 4s$^2$4p$^4$($^3$P)4d   &  $^4$D$  _{5/2}$    & 5.0336 & 5.1603  \\
  5  & 4s$^2$4p$^4$($^3$P)4d   &  $^4$D$  _{3/2}$    & 5.0384 & 5.1688  \\
  6  & 4s$^2$4p$^4$($^3$P)4d   &  $^4$D$  _{7/2}$    & 5.0938 & 5.2190  \\
  7  & 4s$^2$4p$^4$($^3$P)4d   &  $^4$D$  _{1/2}$    & 5.1201 & 5.2515  \\
  8  & 4s$^2$4p$^4$($^3$P)4d   &  $^4$F$  _{9/2}$    & 5.2627 & 5.3963  \\
\hline            								                	 
\end{longtable}   								   					       
			      							   					       

														       
\begin{flushleft}													       
{\small
GRASPa: present calculations from the {\sc grasp} code with 3990 levels\\
GRASPb: earlier calculations of Goyal et al. \cite{ag} from the {\sc grasp} code with 470 levels\\		       
}															       
\end{flushleft} 

\clearpage
\newpage

\renewcommand{\baselinestretch}{1.0}
\footnotesize
\begin{longtable}{@{\extracolsep\fill}rrllll@{}}
\caption{Comparison of oscillator strengths (f) for transitions among the lowest 31 levels of Sn~XVI.  $a{\pm}b \equiv a{\times}$10$^{{\pm}b}$. \\See Table 8 for definition of level indices.}
I      &    J &  GRASPa  &  GRASPb      & FACa      & FACb       \\
 \hline\\
 \endfirsthead\\
\caption[]{(continued)}
I      &    J &  GRASPa  &  GRASPb      & FACa      & FACb       \\
\hline\\
\endhead 
     1 &    3 &    3.604$-$2 &   4.27$-$2  & 3.802$-$2 & 3.839$-$2  \\
     1 &    4 &    6.676$-$4 &   6.51$-$4  & 6.841$-$4 & 6.728$-$4  \\
     1 &    5 &    1.252$-$3 &   9.93$-$4  & 1.307$-$3 & 1.299$-$3  \\
     1 &    7 &    1.809$-$3 &   1.51$-$3  & 1.926$-$3 & 1.916$-$3  \\
     1 &   10 &    4.618$-$3 &   5.45$-$3  & 4.643$-$3 & 4.624$-$3  \\
     1 &   11 &    2.758$-$2 &   2.56$-$2  & 2.925$-$2 & 2.932$-$2  \\
     1 &   12 &    3.143$-$2 &   2.07$-$2  & 3.153$-$2 & 3.182$-$2  \\
     1 &   13 &    2.219$-$3 &   3.07$-$2  & 2.043$-$3 & 2.021$-$3  \\
\hline            								                	 
\end{longtable}   								   					       
			      							   					       

														       
\begin{flushleft}													       
{\small
GRASPa: present calculations with the {\sc grasp} code with 3990 levels \\
GRASPb: calculations of  Goyal et al.  \cite{ag} with the {\sc grasp} code with 470 levels \\
FACa: present calculations with the {\sc fac} code with 3990 levels \\
FACb: present calculations with the {\sc fac} code with 12~137 levels \\
		       
}															       
\end{flushleft} 

\clearpage
\newpage

\renewcommand{\baselinestretch}{1.0}
\footnotesize
\begin{longtable}{@{\extracolsep\fill}rlrllll@{}}
\caption{Comparison of oscillator strengths (f) for the 4s$^2$4p$^4$4d--4s$^2$4p$^4$5p transitions of Sn~XVI.  See Table 8 for definition of all level indices. $a{\pm}b \equiv a{\times}$10$^{{\pm}b}$.}
S. No.  & Level of  4s$^2$4p$^4$4d  & Index   & Level of  4s$^2$4p$^4$5p  & Index  &  Present  & D'Arcy et al. \cite{darcy} \\  \\
 \hline\\
\endfirsthead\\
\caption[]{(continued)}
S. No.  & Level of  4s$^2$4p$^4$4d  & Index   & Level of  4s$^2$4p$^4$5p  & Index  &  Present  & D'Arcy et al. \cite{darcy}  \\  \\
\hline\\
\endhead 
    1 &  $^4$D$  _{7/2 }$     &     6 & $^4$D$^o_{7/2 }$        & 236  &  0.040     &  0.050  \\
    2 &  $^4$F$  _{7/2 }$     &    17 & $^4$D$^o_{5/2 }$        & 247  &  0.104     &  0.131  \\
    3 &  $^2$G$  _{9/2 }$     &    21 & $^2$F$^o_{7/2}$	        & 251  &  0.122     &  0.145  \\
    4 &  $^4$F$  _{9/2 }$     &     8 & $^4$D$^o_{7/2 }$        & 236  &  0.120     &  0.144  \\
    5 &  $^4$F$  _{5/2 }$     &    12 & $^4$D$^o_{3/2 }$        & 242  &  0.094     &  0.130  \\
    6 &  $^4$F$  _{7/2 }$     &    17 & $^4$P$^o_{5/2 }$        & 234  &  1.55--3   &  0.081  \\
    7 &  ($^3$P)$^2$F$_{5/2}$ &    22 & ($^3$P)$^2$D$^o_{3/2}$  & 249  &  0.075	  &  0.067  \\
    8 &  $^2$G$  _{7/2 }$     &    18 & $^2$F$^o_{5/2}$	        & 246  &  0.109	  &  0.115  \\
    9 &  ($^1$D)$^2$F$_{7/2}$ &    24 & ($^1$D)$^2$D$^o_{5/2}$  & 255  &  0.093	  &  0.106  \\
   10 &  $^4$F$  _{3/2 }$     &    14 & ($^3$P)$^2$P$^o_{3/2}$  & 241  &  0.045	  &  0.050  \\
   11 &  $^4$P$  _{5/2 }$     &    15 & ($^3$P)$^2$P$^o_{3/2}$  & 241  &  0.046	  &  0.068  \\
   12 &  ($^3$P)$^2$D$_{5/2}$ &    29 & ($^1$S)$^2$P$^o_{3/2}$  & 278  &  0.049	  &  0.057  \\
   13 &  ($^1$D)$^2$F$_{7/2}$ &    24 & $^4$D$^o_{5/2 }$        & 247  &  0.003	  &  0.056  \\
\\  \hline            								                	 
\end{longtable}   								   					       
			      							   					       

														       
\begin{flushleft}													       
{\small
		       
}															       
\end{flushleft}

\clearpage
\newpage


\TableExplanation

\bigskip
\renewcommand{\arraystretch}{1.0}

\section*{Table 1.\label{tbl1te} Energies (Ryd) for the lowest 375 levels of Tc~IX and their lifetimes ($\tau$, s). }
\begin{tabular}{@{}p{1in}p{6in}@{}}
Index            & Level Index \\
Configuration    & The configuration to which the level belongs \\
Level             & The $LSJ$ designation of the level \\
Energy           & Present energies from the {\sc grasp} code  with 3990 level calculations \\
$\tau$ (s)       & Lifetime of the level in s \\

\end{tabular}
\label{tableII}

\bigskip
\renewcommand{\arraystretch}{1.0}

\section*{Table 2.\label{tbl2te} Energies (Ryd) for the lowest 375 levels of Ru~X and their lifetimes ($\tau$, s). }
\begin{tabular}{@{}p{1in}p{6in}@{}}
Index            & Level Index \\
Configuration    & The configuration to which the level belongs \\
Level             & The $LSJ$ designation of the level \\
Energy           & Present energies from the {\sc grasp} code  with 3990 level calculations \\
$\tau$ (s)       & Lifetime of the level in s \\

\end{tabular}
\label{tableII}

\bigskip
\renewcommand{\arraystretch}{1.0}

\section*{Table 3.\label{tbl3te} Energies (Ryd) for the lowest 375 levels of Rh~XI and their lifetimes ($\tau$, s). }
\begin{tabular}{@{}p{1in}p{6in}@{}}
Index            & Level Index \\
Configuration    & The configuration to which the level belongs \\
Level             & The $LSJ$ designation of the level \\
Energy           & Present energies from the {\sc grasp} code  with 3990 level calculations \\
$\tau$ (s)       & Lifetime of the level in s \\

\end{tabular}
\label{tableII}

\bigskip
\renewcommand{\arraystretch}{1.0}

\section*{Table 4.\label{tbl4te} Energies (Ryd) for the lowest 375 levels of Pd~XII and their lifetimes ($\tau$, s). }
\begin{tabular}{@{}p{1in}p{6in}@{}}
Index            & Level Index \\
Configuration    & The configuration to which the level belongs \\
Level             & The $LSJ$ designation of the level \\
Energy           & Present energies from the {\sc grasp} code  with 3990 level calculations \\
$\tau$ (s)       & Lifetime of the level in s \\

\end{tabular}
\label{tableII}

\bigskip
\renewcommand{\arraystretch}{1.0}

\section*{Table 5.\label{tbl5te} Energies (Ryd) for the lowest 375 levels of Ag~XIII and their lifetimes ($\tau$, s). }
\begin{tabular}{@{}p{1in}p{6in}@{}}
Index            & Level Index \\
Configuration    & The configuration to which the level belongs \\
Level             & The $LSJ$ designation of the level \\
Energy           & Present energies from the {\sc grasp} code  with 3990 level calculations \\
$\tau$ (s)       & Lifetime of the level in s \\

\end{tabular}
\label{tableII}

\bigskip
\renewcommand{\arraystretch}{1.0}

\section*{Table 6.\label{tbl6te} Energies (Ryd) for the lowest 375 levels of Cd~XIV and their lifetimes ($\tau$, s). }
\begin{tabular}{@{}p{1in}p{6in}@{}}
Index            & Level Index \\
Configuration    & The configuration to which the level belongs \\
Level             & The $LSJ$ designation of the level \\
Energy           & Present energies from the {\sc grasp} code  with 3990 level calculations \\
$\tau$ (s)       & Lifetime of the level in s \\

\end{tabular}
\label{tableII}

\bigskip
\renewcommand{\arraystretch}{1.0}

\section*{Table 7.\label{tbl7te} Energies (Ryd) for the lowest 375 levels of In~XV and their lifetimes ($\tau$, s). }
\begin{tabular}{@{}p{1in}p{6in}@{}}
Index            & Level Index \\
Configuration    & The configuration to which the level belongs \\
Level             & The $LSJ$ designation of the level \\
Energy           & Present energies from the {\sc grasp} code  with 3990 level calculations \\
$\tau$ (s)       & Lifetime of the level in s \\

\end{tabular}
\label{tableII}

\bigskip
\renewcommand{\arraystretch}{1.0}

\section*{Table 8.\label{tbl8te} Energies (Ryd) for the lowest 375 levels of Sn~XVI and their lifetimes ($\tau$, s). }
\begin{tabular}{@{}p{1in}p{6in}@{}}
Index            & Level Index \\
Configuration    & The configuration to which the level belongs \\
Level             & The $LSJ$ designation of the level \\
Energy           & Present energies from the {\sc grasp} code  with 3990 level calculations \\
$\tau$ (s)       & Lifetime of the level in s \\

\end{tabular}
\label{tableII}


\bigskip
\section*{Table 9.\label{tbl9te}  Transition wavelengths ($\lambda_{ij}$ in $\rm \AA$), radiative rates (A$_{ji}$ in s$^{-1}$),
 oscillator strengths (f$_{ij}$, dimensionless), and line strengths (S, in atomic units) for electric dipole (E1), and 
A$_{ji}$ for electric quadrupole (E2), magnetic dipole (M1), and magnetic quadrupole (M2) transitions of Tc~IX.  The ratio R(E1) of 
velocity and length forms of A- values for E1 transitions is listed in the last column.}
\begin{tabular}{@{}p{1in}p{6in}@{}}
$i$ and $j$         & The lower ($i$) and upper ($j$) levels of a transition as defined in Table 1.\\
$\lambda_{ij}$      & Transition wavelength (in ${\rm \AA}$) \\
A$^{E1}_{ji}$       & Radiative transition probability (in s$^{-1}$) for the E1 transitions \\
f$^{E1}_{ij}$       & Absorption oscillator strength (dimensionless) for the E1 transitions \\
S$^{E1}$            & Line strength in atomic unit (a.u.), 1 a.u. = 6.460$\times$10$^{-36}$ cm$^2$ esu$^2$ for the E1 transitions \\
A$^{E2}_{ji}$       & Radiative transition probability (in s$^{-1}$) for the E2 transitions \\
A$^{M1}_{ji}$       & Radiative transition probability (in s$^{-1}$) for the M1 transitions \\
A$^{M2}_{ji}$       & Radiative transition probability (in s$^{-1}$) for the M2 transitions \\
R(E1)                     & Ratio of velocity and length forms of A- (or f- and S-) values for the E1 transitions \\
$a{\pm}b$ &  $\equiv a\times{10^{{\pm}b}}$ \\
\end{tabular}
\label{ExplTable9}

\bigskip
\section*{Table 10.\label{tbl10te}  Transition wavelengths ($\lambda_{ij}$ in $\rm \AA$), radiative rates (A$_{ji}$ in s$^{-1}$),
 oscillator strengths (f$_{ij}$, dimensionless), and line strengths (S, in atomic units) for electric dipole (E1), and 
A$_{ji}$ for electric quadrupole (E2), magnetic dipole (M1), and magnetic quadrupole (M2) transitions of Ru~X.  The ratio R(E1) of 
velocity and length forms of A- values for E1 transitions is listed in the last column.}
\begin{tabular}{@{}p{1in}p{6in}@{}}
$i$ and $j$         & The lower ($i$) and upper ($j$) levels of a transition as defined in Table 2.\\
$\lambda_{ij}$      & Transition wavelength (in ${\rm \AA}$) \\
A$^{E1}_{ji}$       & Radiative transition probability (in s$^{-1}$) for the E1 transitions \\
f$^{E1}_{ij}$       & Absorption oscillator strength (dimensionless) for the E1 transitions \\
S$^{E1}$            & Line strength in atomic unit (a.u.), 1 a.u. = 6.460$\times$10$^{-36}$ cm$^2$ esu$^2$ for the E1 transitions \\
A$^{E2}_{ji}$       & Radiative transition probability (in s$^{-1}$) for the E2 transitions \\
A$^{M1}_{ji}$       & Radiative transition probability (in s$^{-1}$) for the M1 transitions \\
A$^{M2}_{ji}$       & Radiative transition probability (in s$^{-1}$) for the M2 transitions \\
R(E1)                     & Ratio of velocity and length forms of A- (or f- and S-) values for the E1 transitions \\
$a{\pm}b$ &  $\equiv a\times{10^{{\pm}b}}$ \\
\end{tabular}
\label{ExplTable10}

\bigskip
\section*{Table 11.\label{tbl11te}  Transition wavelengths ($\lambda_{ij}$ in $\rm \AA$), radiative rates (A$_{ji}$ in s$^{-1}$),
 oscillator strengths (f$_{ij}$, dimensionless), and line strengths (S, in atomic units) for electric dipole (E1), and 
A$_{ji}$ for electric quadrupole (E2), magnetic dipole (M1), and magnetic quadrupole (M2) transitions of Rh~XI.  The ratio R(E1) of 
velocity and length forms of A- values for E1 transitions is listed in the last column.}
\begin{tabular}{@{}p{1in}p{6in}@{}}
$i$ and $j$         & The lower ($i$) and upper ($j$) levels of a transition as defined in Table 3.\\
$\lambda_{ij}$      & Transition wavelength (in ${\rm \AA}$) \\
A$^{E1}_{ji}$       & Radiative transition probability (in s$^{-1}$) for the E1 transitions \\
f$^{E1}_{ij}$       & Absorption oscillator strength (dimensionless) for the E1 transitions \\
S$^{E1}$            & Line strength in atomic unit (a.u.), 1 a.u. = 6.460$\times$10$^{-36}$ cm$^2$ esu$^2$ for the E1 transitions \\
A$^{E2}_{ji}$       & Radiative transition probability (in s$^{-1}$) for the E2 transitions \\
A$^{M1}_{ji}$       & Radiative transition probability (in s$^{-1}$) for the M1 transitions \\
A$^{M2}_{ji}$       & Radiative transition probability (in s$^{-1}$) for the M2 transitions \\
R(E1)                     & Ratio of velocity and length forms of A- (or f- and S-) values for the E1 transitions \\
$a{\pm}b$ &  $\equiv a\times{10^{{\pm}b}}$ \\
\end{tabular}
\label{ExplTable11}

\bigskip
\section*{Table 12.\label{tbl12te}  Transition wavelengths ($\lambda_{ij}$ in $\rm \AA$), radiative rates (A$_{ji}$ in s$^{-1}$),
 oscillator strengths (f$_{ij}$, dimensionless), and line strengths (S, in atomic units) for electric dipole (E1), and 
A$_{ji}$ for electric quadrupole (E2), magnetic dipole (M1), and magnetic quadrupole (M2) transitions of Pd~XII.  The ratio R(E1) of 
velocity and length forms of A- values for E1 transitions is listed in the last column.}
\begin{tabular}{@{}p{1in}p{6in}@{}}
$i$ and $j$         & The lower ($i$) and upper ($j$) levels of a transition as defined in Table 4.\\
$\lambda_{ij}$      & Transition wavelength (in ${\rm \AA}$) \\
A$^{E1}_{ji}$       & Radiative transition probability (in s$^{-1}$) for the E1 transitions \\
f$^{E1}_{ij}$       & Absorption oscillator strength (dimensionless) for the E1 transitions \\
S$^{E1}$            & Line strength in atomic unit (a.u.), 1 a.u. = 6.460$\times$10$^{-36}$ cm$^2$ esu$^2$ for the E1 transitions \\
A$^{E2}_{ji}$       & Radiative transition probability (in s$^{-1}$) for the E2 transitions \\
A$^{M1}_{ji}$       & Radiative transition probability (in s$^{-1}$) for the M1 transitions \\
A$^{M2}_{ji}$       & Radiative transition probability (in s$^{-1}$) for the M2 transitions \\
R(E1)                     & Ratio of velocity and length forms of A- (or f- and S-) values for the E1 transitions \\
$a{\pm}b$ &  $\equiv a\times{10^{{\pm}b}}$ \\
\end{tabular}
\label{ExplTable12}

\bigskip
\section*{Table 13.\label{tbl13te}  Transition wavelengths ($\lambda_{ij}$ in $\rm \AA$), radiative rates (A$_{ji}$ in s$^{-1}$),
 oscillator strengths (f$_{ij}$, dimensionless), and line strengths (S, in atomic units) for electric dipole (E1), and 
A$_{ji}$ for electric quadrupole (E2), magnetic dipole (M1), and magnetic quadrupole (M2) transitions of Ag~XIII.  The ratio R(E1) of 
velocity and length forms of A- values for E1 transitions is listed in the last column.}
\begin{tabular}{@{}p{1in}p{6in}@{}}
$i$ and $j$         & The lower ($i$) and upper ($j$) levels of a transition as defined in Table 5.\\
$\lambda_{ij}$      & Transition wavelength (in ${\rm \AA}$) \\
A$^{E1}_{ji}$       & Radiative transition probability (in s$^{-1}$) for the E1 transitions \\
f$^{E1}_{ij}$       & Absorption oscillator strength (dimensionless) for the E1 transitions \\
S$^{E1}$            & Line strength in atomic unit (a.u.), 1 a.u. = 6.460$\times$10$^{-36}$ cm$^2$ esu$^2$ for the E1 transitions \\
A$^{E2}_{ji}$       & Radiative transition probability (in s$^{-1}$) for the E2 transitions \\
A$^{M1}_{ji}$       & Radiative transition probability (in s$^{-1}$) for the M1 transitions \\
A$^{M2}_{ji}$       & Radiative transition probability (in s$^{-1}$) for the M2 transitions \\
R(E1)                     & Ratio of velocity and length forms of A- (or f- and S-) values for the E1 transitions \\
$a{\pm}b$ &  $\equiv a\times{10^{{\pm}b}}$ \\
\end{tabular}
\label{ExplTable13}

\bigskip
\section*{Table 14.\label{tbl14te}  Transition wavelengths ($\lambda_{ij}$ in $\rm \AA$), radiative rates (A$_{ji}$ in s$^{-1}$),
 oscillator strengths (f$_{ij}$, dimensionless), and line strengths (S, in atomic units) for electric dipole (E1), and 
A$_{ji}$ for electric quadrupole (E2), magnetic dipole (M1), and magnetic quadrupole (M2) transitions of Cd~XIV.  The ratio R(E1) of 
velocity and length forms of A- values for E1 transitions is listed in the last column.}
\begin{tabular}{@{}p{1in}p{6in}@{}}
$i$ and $j$         & The lower ($i$) and upper ($j$) levels of a transition as defined in Table 6.\\
$\lambda_{ij}$      & Transition wavelength (in ${\rm \AA}$) \\
A$^{E1}_{ji}$       & Radiative transition probability (in s$^{-1}$) for the E1 transitions \\
f$^{E1}_{ij}$       & Absorption oscillator strength (dimensionless) for the E1 transitions \\
S$^{E1}$            & Line strength in atomic unit (a.u.), 1 a.u. = 6.460$\times$10$^{-36}$ cm$^2$ esu$^2$ for the E1 transitions \\
A$^{E2}_{ji}$       & Radiative transition probability (in s$^{-1}$) for the E2 transitions \\
A$^{M1}_{ji}$       & Radiative transition probability (in s$^{-1}$) for the M1 transitions \\
A$^{M2}_{ji}$       & Radiative transition probability (in s$^{-1}$) for the M2 transitions \\
R(E1)                     & Ratio of velocity and length forms of A- (or f- and S-) values for the E1 transitions \\
$a{\pm}b$ &  $\equiv a\times{10^{{\pm}b}}$ \\
\end{tabular}
\label{ExplTable14}

\bigskip
\section*{Table 15.\label{tbl15te}  Transition wavelengths ($\lambda_{ij}$ in $\rm \AA$), radiative rates (A$_{ji}$ in s$^{-1}$),
 oscillator strengths (f$_{ij}$, dimensionless), and line strengths (S, in atomic units) for electric dipole (E1), and 
A$_{ji}$ for electric quadrupole (E2), magnetic dipole (M1), and magnetic quadrupole (M2) transitions of In~XV.  The ratio R(E1) of 
velocity and length forms of A- values for E1 transitions is listed in the last column.}
\begin{tabular}{@{}p{1in}p{6in}@{}}
$i$ and $j$         & The lower ($i$) and upper ($j$) levels of a transition as defined in Table 7.\\
$\lambda_{ij}$      & Transition wavelength (in ${\rm \AA}$) \\
A$^{E1}_{ji}$       & Radiative transition probability (in s$^{-1}$) for the E1 transitions \\
f$^{E1}_{ij}$       & Absorption oscillator strength (dimensionless) for the E1 transitions \\
S$^{E1}$            & Line strength in atomic unit (a.u.), 1 a.u. = 6.460$\times$10$^{-36}$ cm$^2$ esu$^2$ for the E1 transitions \\
A$^{E2}_{ji}$       & Radiative transition probability (in s$^{-1}$) for the E2 transitions \\
A$^{M1}_{ji}$       & Radiative transition probability (in s$^{-1}$) for the M1 transitions \\
A$^{M2}_{ji}$       & Radiative transition probability (in s$^{-1}$) for the M2 transitions \\
R(E1)                     & Ratio of velocity and length forms of A- (or f- and S-) values for the E1 transitions \\
$a{\pm}b$ &  $\equiv a\times{10^{{\pm}b}}$ \\
\end{tabular}
\label{ExplTable15}

\bigskip
\section*{Table 16.\label{tbl16te}  Transition wavelengths ($\lambda_{ij}$ in $\rm \AA$), radiative rates (A$_{ji}$ in s$^{-1}$),
 oscillator strengths (f$_{ij}$, dimensionless), and line strengths (S, in atomic units) for electric dipole (E1), and 
A$_{ji}$ for electric quadrupole (E2), magnetic dipole (M1), and magnetic quadrupole (M2) transitions of Sn~XVI.  The ratio R(E1) of 
velocity and length forms of A- values for E1 transitions is listed in the last column.}
\begin{tabular}{@{}p{1in}p{6in}@{}}
$i$ and $j$         & The lower ($i$) and upper ($j$) levels of a transition as defined in Table 8.\\
$\lambda_{ij}$      & Transition wavelength (in ${\rm \AA}$) \\
A$^{E1}_{ji}$       & Radiative transition probability (in s$^{-1}$) for the E1 transitions \\
f$^{E1}_{ij}$       & Absorption oscillator strength (dimensionless) for the E1 transitions \\
S$^{E1}$            & Line strength in atomic unit (a.u.), 1 a.u. = 6.460$\times$10$^{-36}$ cm$^2$ esu$^2$ for the E1 transitions \\
A$^{E2}_{ji}$       & Radiative transition probability (in s$^{-1}$) for the E2 transitions \\
A$^{M1}_{ji}$       & Radiative transition probability (in s$^{-1}$) for the M1 transitions \\
A$^{M2}_{ji}$       & Radiative transition probability (in s$^{-1}$) for the M2 transitions \\
R(E1)                     & Ratio of velocity and length forms of A- (or f- and S-) values for the E1 transitions \\
$a{\pm}b$ &  $\equiv a\times{10^{{\pm}b}}$ \\
\end{tabular}
\label{ExplTable16}

\datatables 



\setlength{\LTleft}{0pt}
\setlength{\LTright}{0pt} 


\setlength{\tabcolsep}{0.5\tabcolsep}

\renewcommand{\arraystretch}{1.0}

\footnotesize 

\begin{longtable}{@{\extracolsep\fill}rllll@{}}
\caption{Energies (Ryd) for the lowest 375 levels of Tc~IX and their lifetimes ($\tau$, s). $a{\pm}b \equiv a{\times}$10$^{{\pm}b}$. See page\ \pageref{tbl1te} for Explanation of Tables}
Index  & Configuration                       & Level              & Energy  & $\tau$  (s)   \\
\hline\\
\endfirsthead\\
\caption[]{(continued)}
Index  & Configuration                       & Level              & Energy  & $\tau$  (s)   \\
\hline\\
\endhead
    1   & 4s$^2$4p$^5$                       &   $^2$P$^o_{3/2 }$    &  0.00000    &    .......                \\
    2   & 4s$^2$4p$^5$                       &   $^2$P$^o_{1/2 }$    &  0.24920    &    2.719$-$03     \\
    3   & 4s4p$^6$                           &   $^2$S$  _{1/2 }$    &  2.47330    &    2.497$-$10     \\
    4   & 4s$^2$4p$^4$($^3$P)4d              &   $^4$D$  _{5/2 }$    &  3.16023    &    2.333$-$08     \\
    5   & 4s$^2$4p$^4$($^3$P)4d              &   $^4$D$  _{7/2 }$    &  3.17167    &    1.248$-$02     \\
    6   & 4s$^2$4p$^4$($^3$P)4d              &   $^4$D$  _{3/2 }$    &  3.17509    &    2.214$-$08     \\
    7   & 4s$^2$4p$^4$($^3$P)4d              &   $^4$D$  _{1/2 }$    &  3.20934    &    1.375$-$08     \\
    8   & 4s$^2$4p$^4$($^3$P)4d              &   $^4$F$  _{9/2 }$    &  3.31129    &    1.202$-$01     \\
    9   & 4s$^2$4p$^4$($^3$P)4d              &   $^4$F$  _{7/2 }$    &  3.37669    &    2.869$-$02     \\
   10   & 4s$^2$4p$^4$($^1$D)4d              &   $^2$P$  _{1/2 }$    &  3.42622    &    5.773$-$09     \\
   ............  \\
   .............  \\
   \hline  											      
			      							   					       

\\
\end{longtable}


\begin{longtable}{@{\extracolsep\fill}rllll@{}}
\caption{Energies (Ryd) for the lowest 375 levels of Ru~X and their lifetimes ($\tau$, s). $a{\pm}b \equiv a{\times}$10$^{{\pm}b}$. See page\ \pageref{tbl2te} for Explanation of Tables}
Index  & Configuration                       & Level              & Energy  & $\tau$  (s)   \\
\hline\\
\endfirsthead\\
\caption[]{(continued)}
Index  & Configuration                       & Level              & Energy  & $\tau$  (s)   \\
\hline\\
\endhead
    1  & 4s$^2$4p$^5$			   &   $^2$P$^o_{3/2 }$    &	      0.00000  &  ..........	 \\
    2  & 4s$^2$4p$^5$			   &   $^2$P$^o_{1/2 }$    &	      0.29593  &  1.623$-$03	 \\
    3  & 4s4p$^6$			   &   $^2$S$  _{1/2 }$    &	      2.68709  &  1.984$-$10	 \\
    4  & 4s$^2$4p$^4$($^3$P)4d  	   &   $^4$D$  _{5/2 }$    &	      3.43132  &  1.934$-$08	 \\
    5  & 4s$^2$4p$^4$($^3$P)4d  	   &   $^4$D$  _{3/2 }$    &	      3.44654  &  1.550$-$08	 \\
    6  & 4s$^2$4p$^4$($^3$P)4d  	   &   $^4$D$  _{7/2 }$    &	      3.44741  &  9.295$-$03	 \\
    7  & 4s$^2$4p$^4$($^3$P)4d  	   &   $^4$D$  _{1/2 }$    &	      3.48712  &  8.664$-$09	 \\
    8  & 4s$^2$4p$^4$($^3$P)4d  	   &   $^4$F$  _{9/2 }$    &	      3.59149  &  9.447$-$02	 \\
    9  & 4s$^2$4p$^4$($^3$P)4d  	   &   $^4$F$  _{7/2 }$    &	      3.66026  &  2.152$-$02	 \\
   10  & 4s$^2$4p$^4$($^3$P)4d  	   &   $^4$F$  _{3/2 }$    &	      3.72684  &  2.333$-$09	 \\
  ............  \\
   .............  \\
\hline  											      
			      							   					       

\\
\end{longtable}

\clearpage
\newpage

\begin{longtable}{@{\extracolsep\fill}rllll@{}}
\caption{Energies (Ryd) for the lowest 375 levels of Rh~XI and their lifetimes ($\tau$, s). $a{\pm}b \equiv a{\times}$10$^{{\pm}b}$. See page\ \pageref{tbl3te} for Explanation of Tables}
Index  & Configuration                       & Level              & Energy  & $\tau$  (s)   \\
\hline\\
\endfirsthead\\
\caption[]{(continued)}
Index  & Configuration                       & Level              & Energy  & $\tau$  (s)   \\
\hline\\
\endhead
    1  &  4s$^2$4p$^5$  		     &   $^2$P$^o_{3/2 }$    &  0.00000  &  .........	\\
    2  &  4s$^2$4p$^5$  		     &   $^2$P$^o_{1/2 }$    &  0.34817  &  9.967$-$04   \\
    3  &  4s4p$^6$			     &   $^2$S$  _{1/2 }$    &  2.90522  &  1.619$-$10   \\
    4  &  4s$^2$4p$^4$($^3$P)4d 	     &   $^4$D$  _{5/2 }$    &  3.70088  &  1.647$-$08   \\
    5  &  4s$^2$4p$^4$($^3$P)4d 	     &   $^4$D$  _{3/2 }$    &  3.71591  &  1.132$-$08   \\
    6  &  4s$^2$4p$^4$($^3$P)4d 	     &   $^4$D$  _{7/2 }$    &  3.72237  &  7.131$-$03   \\
    7  &  4s$^2$4p$^4$($^3$P)4d 	     &   $^4$D$  _{1/2 }$    &  3.76303  &  5.769$-$09   \\
    8  &  4s$^2$4p$^4$($^3$P)4d 	     &   $^4$F$  _{9/2 }$    &  3.87054  &  7.600$-$02   \\
    9  &  4s$^2$4p$^4$($^3$P)4d 	     &   $^4$F$  _{7/2 }$    &  3.94111  &  1.650$-$02   \\
   10  &  4s$^2$4p$^4$($^3$P)4d 	     &   $^4$F$  _{3/2 }$    &  4.01354  &  1.686$-$09   \\
  ............  \\
   .............  \\
\hline  											      
			      							   					       

\\
\end{longtable}


\begin{longtable}{@{\extracolsep\fill}rllll@{}}
\caption{Energies (Ryd) for the lowest 375 levels of Pd~XII and their lifetimes ($\tau$, s). $a{\pm}b \equiv a{\times}$10$^{{\pm}b}$. See page\ \pageref{tbl4te} for Explanation of Tables}
Index  & Configuration                       & Level              & Energy  & $\tau$  (s)   \\
\hline\\
\endfirsthead\\
\caption[]{(continued)}
Index  & Configuration                       & Level              & Energy  & $\tau$  (s)   \\
\hline\\
\endhead
    1   &   4s$^2$4p$^5$		       &   $^2$P$^o_{3/2 }$    &   0.00000    &    ..........   \\
    2   &   4s$^2$4p$^5$		       &   $^2$P$^o_{1/2 }$    &   0.40633    &    6.269$-$04   \\
    3   &   4s4p$^6$			       &   $^2$S$  _{1/2 }$    &   3.12830    &    1.350$-$10   \\
    4   &   4s$^2$4p$^4$($^3$P)4d	       &   $^4$D$  _{5/2 }$    &   3.96918    &    1.441$-$08   \\
    5   &   4s$^2$4p$^4$($^3$P)4d	       &   $^4$D$  _{3/2 }$    &   3.98343    &    8.570$-$09   \\
    6   &   4s$^2$4p$^4$($^3$P)4d	       &   $^4$D$  _{7/2 }$    &   3.99685    &    5.613$-$03   \\
    7   &   4s$^2$4p$^4$($^3$P)4d	       &   $^4$D$  _{1/2 }$    &   4.03720    &    4.012$-$09   \\
    8   &   4s$^2$4p$^4$($^3$P)4d	       &   $^4$F$  _{9/2 }$    &   4.14889    &    6.218$-$02   \\
    9   &   4s$^2$4p$^4$($^3$P)4d	       &   $^2$F$  _{7/2 }$    &   4.21970    &    1.281$-$02   \\
   10   &   4s$^2$4p$^4$($^3$P)4d	       &   $^4$F$  _{3/2 }$    &   4.29833    &    1.321$-$09   \\
  ............  \\
   .............  \\
\hline  											      
			      							   					       

\\
\end{longtable}

\clearpage
\newpage

\begin{longtable}{@{\extracolsep\fill}rllll@{}}
\caption{Energies (Ryd) for the lowest 375 levels of Ag~XIII and their lifetimes ($\tau$, s). $a{\pm}b \equiv a{\times}$10$^{{\pm}b}$. See page\ \pageref{tbl5te} for Explanation of Tables}
Index  & Configuration                       & Level              & Energy  & $\tau$  (s)   \\
\hline\\
\endfirsthead\\
\caption[]{(continued)}
Index  & Configuration                       & Level              & Energy  & $\tau$  (s)   \\
\hline\\
\endhead
    1   &   4s$^2$4p$^5$		       &   $^2$P$^o_{3/2 }$    &   0.00000   & ..........    \\  
    2   &   4s$^2$4p$^5$		       &   $^2$P$^o_{1/2 }$    &   0.47082   & 4.029$-$04    \\  
    3   &   4s4p$^6$			       &   $^2$S$  _{1/2 }$    &   3.35690   & 1.146$-$10    \\
    4   &   4s$^2$4p$^4$($^3$P)4d	       &   $^4$D$  _{5/2 }$    &   4.23644   & 1.294$-$08    \\
    5   &   4s$^2$4p$^4$($^3$P)4d	       &   $^4$D$  _{3/2 }$    &   4.24927   & 6.686$-$09    \\
    6   &   4s$^2$4p$^4$($^3$P)4d	       &   $^4$D$  _{7/2 }$    &   4.27108   & 4.518$-$03    \\
    7   &   4s$^2$4p$^4$($^3$P)4d	       &   $^4$D$  _{1/2 }$    &   4.30979   & 2.885$-$09    \\
    8   &   4s$^2$4p$^4$($^3$P)4d	       &   $^4$F$  _{9/2 }$    &   4.42694   & 5.144$-$02    \\
    9   &   4s$^2$4p$^4$($^3$P)4d	       &   $^2$F$  _{7/2 }$    &   4.49649   & 9.992$-$03    \\
   10   &   4s$^2$4p$^4$($^1$D)4d	       &   $^2$P$  _{3/2 }$    &   4.58162   & 1.096$-$09    \\
  ............  \\
   .............  \\
\hline  											      
			      							   					       

\\
\end{longtable}


\begin{longtable}{@{\extracolsep\fill}rllll@{}}
\caption{Energies (Ryd) for the lowest 375 levels of Cd~XIV and their lifetimes ($\tau$, s). $a{\pm}b \equiv a{\times}$10$^{{\pm}b}$. See page\ \pageref{tbl6te} for Explanation of Tables}
Index  & Configuration                       & Level              & Energy  & $\tau$  (s)   \\
\hline\\
\endfirsthead\\
\caption[]{(continued)}
Index  & Configuration                       & Level              & Energy  & $\tau$  (s)   \\
\hline\\
\endhead
    1   &   4s$^2$4p$^5$		       &   $^2$P$^o_{3/2 }$    &  0.00000   &	..........     \\
    2   &   4s$^2$4p$^5$		       &   $^2$P$^o_{1/2 }$    &  0.54211   &	2.639$-$04     \\
    3   &   4s4p$^6$			       &   $^2$S$  _{1/2 }$    &  3.59158   &	9.870$-$11     \\
    4   &   4s$^2$4p$^4$($^3$P)4d	       &   $^4$D$  _{5/2 }$    &  4.50283   &	1.194$-$08     \\
    5   &   4s$^2$4p$^4$($^3$P)4d	       &   $^4$D$  _{3/2 }$    &  4.51361   &	5.352$-$09     \\
    6   &   4s$^2$4p$^4$($^3$P)4d	       &   $^4$D$  _{7/2 }$    &  4.54523   &	3.711$-$03     \\
    7   &   4s$^2$4p$^4$($^3$P)4d	       &   $^4$D$  _{1/2 }$    &  4.58098   &	2.128$-$09     \\
    8   &   4s$^2$4p$^4$($^3$P)4d	       &   $^4$F$  _{9/2 }$    &  4.70503   &	4.280$-$02     \\
    9   &   4s$^2$4p$^4$($^3$P)4d	       &   $^2$F$  _{7/2 }$    &  4.77190   &	7.798$-$03     \\
   10   &   4s$^2$4p$^4$($^1$S)4d	       &   $^2$D$  _{3/2 }$    &  4.86371   &	9.504$-$10     \\
  ............  \\
   .............  \\
\hline  											      
			      							   					       

\\
\end{longtable}

\clearpage
\newpage

\begin{longtable}{@{\extracolsep\fill}rllll@{}}
\caption{Energies (Ryd) for the lowest 375 levels of In~XV and their lifetimes ($\tau$, s). $a{\pm}b \equiv a{\times}$10$^{{\pm}b}$. See page\ \pageref{tbl7te} for Explanation of Tables}
Index  & Configuration                       & Level              & Energy  & $\tau$  (s)   \\
\hline\\
\endfirsthead\\
\caption[]{(continued)}
Index  & Configuration                       & Level              & Energy  & $\tau$  (s)   \\
\hline\\
\endhead
    1  & 4s$^2$4p$^5$			    &	$^2$P$^o_{3/2 }$    &	 0.00000   & ..........  \\
    2  & 4s$^2$4p$^5$			    &	$^2$P$^o_{1/2 }$    &	 0.62063   & 1.758$-$04  \\
    3  & 4s4p$^6$			    &	$^2$S$  _{1/2 }$    &	 3.83286   & 8.617$-$11  \\
    4  & 4s$^2$4p$^4$($^3$P)4d  	    &	$^4$D$  _{5/2 }$    &	 4.76852   & 1.132$-$08  \\
    5  & 4s$^2$4p$^4$($^3$P)4d  	    &	$^4$D$  _{3/2 }$    &	 4.77661   & 4.376$-$09  \\
    6  & 4s$^2$4p$^4$($^3$P)4d  	    &	$^4$D$  _{7/2 }$    &	 4.81945   & 3.107$-$03  \\
    7  & 4s$^2$4p$^4$($^1$D)4d  	    &	$^2$P$  _{1/2 }$    &	 4.85100   & 1.600$-$09  \\
    8  & 4s$^2$4p$^4$($^3$P)4d  	    &	$^4$F$  _{9/2 }$    &	 4.98351   & 3.564$-$02  \\
    9  & 4s$^2$4p$^4$($^3$P)4d  	    &	$^2$F$  _{7/2 }$    &	 5.04632   & 6.074$-$03  \\
   10  & 4s$^2$4p$^4$($^1$S)4d  	    &	$^2$D$  _{3/2 }$    &	 5.14484   & 8.535$-$10  \\
  ............  \\
   .............  \\
\hline  											      
			      							   					       

\\
\end{longtable}


\begin{longtable}{@{\extracolsep\fill}rllll@{}}
\caption{Energies (Ryd) for the lowest 375 levels of Sn~XVI and their lifetimes ($\tau$, s). $a{\pm}b \equiv a{\times}$10$^{{\pm}b}$. See page\ \pageref{tbl8te} for Explanation of Tables}
Index  & Configuration                       & Level              & Energy  & $\tau$  (s)   \\
\hline\\
\endfirsthead\\
\caption[]{(continued)}
Index  & Configuration                       & Level              & Energy  & $\tau$  (s)   \\
\hline\\
\endhead
    1   &   4s$^2$4p$^5$		       &   $^2$P$^o_{3/2 }$    &   0.00000   & ..........   \\
    2   &   4s$^2$4p$^5$		       &   $^2$P$^o_{1/2 }$    &   0.70689   & 1.190$-$04   \\
    3   &   4s4p$^6$			       &   $^2$S$  _{1/2 }$    &   4.08124   & 7.612$-$11   \\
    4   &   4s$^2$4p$^4$($^3$P)4d	       &   $^4$D$  _{5/2 }$    &   5.03362   & 1.104$-$08   \\
    5   &   4s$^2$4p$^4$($^3$P)4d	       &   $^4$D$  _{3/2 }$    &   5.03844   & 3.644$-$09   \\
    6   &   4s$^2$4p$^4$($^3$P)4d	       &   $^4$D$  _{7/2 }$    &   5.09383   & 2.648$-$03   \\
    7   &   4s$^2$4p$^4$($^3$P)4d	       &   $^4$D$  _{1/2 }$    &   5.12007   & 1.220$-$09   \\
    8   &   4s$^2$4p$^4$($^3$P)4d	       &   $^4$F$  _{9/2 }$    &   5.26266   & 2.957$-$02   \\
    9   &   4s$^2$4p$^4$($^3$P)4d	       &   $^2$F$  _{7/2 }$    &   5.32015   & 4.715$-$03   \\
   10   &   4s$^2$4p$^4$($^1$S)4d	       &   $^2$D$  _{3/2 }$    &   5.42518   & 7.894$-$10   \\
  ............  \\
   .............  \\
\hline  											      
			      							   					       

\\
\end{longtable}

\end{document}